\documentclass{elsart}

\usepackage{graphicx}

\usepackage{amssymb}

\usepackage[english]{babel}

\def\Neqfour{\mathcal{N}=4}
\def\Neqone{\mathcal{N}=1}
\def\Nc{N_c}
\def\SUNc{{\rm SU}(\Nc)}
\def\e{\epsilon}

\def\del{\partial}
\def\Tr{\, {\rm Tr}}

\def\spa#1.#2{\left\langle#1\,#2\right\rangle}
\def\spb#1.#2{\left[#1\,#2\right]}
\def\tree{{(0)}}

\def\oneloop{{(1)}}
\def\twoloop{{(2)}}

\def\Ord{\mathcal{O}}
\def\cA{\mathcal{A}}

\def\bI{\mathbf{I}}
\def\bT{\mathbf{T}}
\def\bH{\mathbf{H}}
\def\lr{\leftrightarrow}
\def\Li{\mathop{\rm Li}\nolimits}
\def\ggtogg{{gg \to gg}}
\def\bi{\begin{itemize}}
\def\ei{\end{itemize}}
\def\ben{\begin{enumerate}}
\def\een{\end{enumerate}}
\def\be{\begin{equation}}
\def\ee{\end{equation}}
\def\bea{\begin{eqnarray}}
\def\eea{\end{eqnarray}}
\def\fig#1{fig.~{\ref{#1}}}

\def\eqn#1{eq.~(\ref{#1})}

\renewcommand{\theequation}{\arabic{equation}}
\begin{document}

\begin{frontmatter}



\selectlanguage{english}
\title{ ~~ \\ ~~ \\ ~~ \\
 $\Neqfour$ Super-Yang-Mills Theory, \\
 QCD and Collider Physics}

\vspace{0.6cm}

\selectlanguage{english}
\author[authorlabel1]{ Z. Bern}
\author[authorlabel2]{ L.J. Dixon}\footnote{Presented at {\it Strings 2004}}
\author[authorlabel3]{ D.A. Kosower}

\address[authorlabel1]{Department of Physics \& Astronomy, UCLA,
	Los Angeles, CA 90095-1547, USA}
\address[authorlabel2]{SLAC, Stanford University, Stanford, CA 94309, USA, and \\
        IPPP, University of Durham, Durham DH1 3LE, England}
\address[authorlabel3]{Service de Physique Th\'eorique, CEA--Saclay,
	F-91191 Gif-sur-Yvette cedex, France}

\begin{abstract}
We review how (dimensionally regulated) scattering amplitudes in 
$\Neqfour$ super-Yang-Mills theory provide a useful testing
ground for perturbative QCD calculations relevant to collider physics,
as well as another avenue for investigating the AdS/CFT correspondence.
We describe the iterative relation for two-loop scattering amplitudes
in $\Neqfour$ super-Yang-Mills theory found in {\it C. Anastasiou et al., 
Phys. Rev. Lett. 91:251602 (2003),} and discuss recent progress toward
extending it to three loops.
%
%
%
\end{abstract}
\end{frontmatter}


\selectlanguage{english}


\section{Introduction and Collider Physics Motivation}
\label{IntroSection}

Maximally supersymmetric ($\Neqfour$) Yang-Mills theory (MSYM)
is unique in many ways.  Its properties are uniquely specified by
the gauge group, say $\SUNc$, and the value of the gauge coupling $g$.  
It is conformally invariant for any value of $g$.
Although gravity is not present in its usual formulation, 
MSYM is connected to gravity and string theory through the AdS/CFT 
correspondence~\cite{AdSCFT}.  Because this correspondence is
a weak-strong coupling duality, it is difficult to verify
quantitatively for general observables.  On the other hand, 
such checks are possible and have been remarkably
successful for quantities protected by supersymmetry such as BPS
operators~\cite{MaldacenaChecks}, 
or when an additional expansion parameter is available, 
such as the number of fields in sequences of composite, large $R$-charge
operators~\cite{BMN,MZ,BKS,BFST,RT,BDS}.

It is interesting to study even more observables in perturbative MSYM,
in order to see how the simplicity of the strong coupling limit
is reflected in the structure of the weak coupling expansion.
The strong coupling limit should be even simpler when the large-$\Nc$
limit is taken simultaneously, as it corresponds to a weakly-coupled
supergravity theory in a background with a large radius of curvature.
There are different ways to study perturbative MSYM.  One approach
is via computation of the anomalous dimensions of composite, gauge
invariant operators~\cite{AdSCFT,BMN,MZ,BKS,BFST,RT,BDS}.
Another possibility~\cite{PlanarMSYM}, discussed here,
is to study the scattering amplitudes for (regulated) plane-wave 
elementary field excitations such as gluons and gluinos.  

One of the virtues of the latter approach is that perturbative 
MSYM scattering amplitudes share many qualitative properties with 
QCD amplitudes in the regime probed at high-energy colliders.  
Yet the results and the computations (when organized in the right way)
are typically significantly simpler.  In this way, MSYM serves as a 
testing ground for many aspects of perturbative QCD.
MSYM loop amplitudes can be considered as components of QCD loop
amplitudes.  Depending on one's point of view, they can be considered
either ``the simplest pieces'' (in terms of the rank of the loop momentum
tensors in the numerator of the amplitude)~\cite{LoopReview,BRY},
or ``the most complicated pieces'' in terms of the degree of
transcendentality (see section~\ref{AnomDimSection})
of the special functions entering the final
results~\cite{KLOV}.  As discussed in section~\ref{AnomDimSection},
the latter interpretation links recent three-loop
anomalous dimension results in QCD~\cite{MVVNNLO} 
to those in the spin-chain approach to MSYM~\cite{BKS}.

The most direct experimental probes of short-distance physics are
collider experiments at the energy frontier.  For the next decade,
that frontier is at hadron colliders --- Run II of the Fermilab Tevatron
now, followed by startup of the CERN Large Hadron Collider in 2007.
New physics at colliders always contends with Standard Model backgrounds.  
At hadron colliders, {\it all} physics processes --- signals and backgrounds ---
are inherently QCD processes.  Hence it is important to be able to predict
them theoretically as precisely as possible.   The cross section for a 
``hard,'' or short-distance-dominated processes, can be
factorized~\cite{Factorization} into a partonic cross section, which can be
computed order by order in perturbative QCD, convoluted with nonperturbative
but {\it measurable} parton distribution functions (pdfs).  
For example, the cross section for producing a pair of jets 
(plus anything else) in a $p\bar{p}$ collision is given by
\bea
 \sigma_{p\bar{p} \to jjX}(s) &=& \sum_{a,b} \int_0^1 dx_1 dx_2
 \ f_a(x_1;\mu_F) \bar{f}_b(x_2;\mu_F)
\nonumber \\
&& \hskip1cm \times
   \hat{\sigma}_{ab \to jjX}(sx_1x_2;\mu_F,\mu_R;\alpha_s(\mu_R)),
\eea
where $s$ is the squared center-of-mass energy, $x_{1,2}$ are the
longitudinal (light-cone) fractions of the $p,\bar{p}$ momentum
carried by partons $a,b$, which may be quarks, anti-quarks or gluons.
The experimental definition of a jet is an involved one which need not
concern us here.  The pdf $f_a(x,\mu_F)$ gives the probability for 
finding parton $a$ with momentum fraction $x$ inside the proton; 
similarly $\bar{f}_b$ is the probability for finding parton $b$
in the antiproton. The pdfs depend logarithmically
on the factorization scale $\mu_F$, or transverse resolution with which
the proton is examined.  The Mellin moments of $f_a(x,\mu_F)$ are forward 
matrix elements of leading-twist operators in the proton, renormalized at the
scale $\mu_F$.  The quark distribution function $q(x,\mu)$, for example, 
obeys $\int_0^1 dx \, x^j \, q(x,\mu) 
= \langle p | [ \bar{q} \gamma^+ \del_+^j q ]_{(\mu)} | p \rangle$.

\section{Ingredients for a NNLO Calculation}
\label{IngredientsSection}

Many hadron collider measurements can benefit from predictions
that are accurate to next-to-next-to-leading order (NNLO) in QCD.
Three separate ingredients enter such an NNLO computation; 
only the third depends on the process:
\begin{enumerate}
\item The experimental value of the QCD coupling 
$\alpha_s(\mu_R)$ must be
determined at one value of the renormalization scale $\mu_R$ (for example
$m_Z$), and its evolution in $\mu_R$ computed using the 3-loop 
$\beta$-function, which has been known since 1980~\cite{MSbarb2}.
\item The experimental values for the pdfs $f_a(x,\mu_F)$
must be determined, ideally using predictions at the NNLO level, 
as are available for deep-inelastic scattering~\cite{NNLODIS}
and more recently Drell-Yan production~\cite{NNLODY}.  The evolution
of pdfs in $\mu_F$ to NNLO accuracy has very recently been completed,
after a multi-year effort by Moch, Vermaseren and Vogt~\cite{MVVNNLO}
(previously, approximations to the NNLO kernel were
available~\cite{NNLOPDFApprox}).
\item The NNLO terms in the expansion of the partonic cross sections
must be computed for the hadronic process in question.  For example, 
the parton cross sections for jet production has the expansion,
\be
 \hat{\sigma}_{ab \to jjX} 
 = \alpha_s^2 ( A + \alpha_s B + \alpha_s^2 C + \ldots ).
\label{NNLOjj}
\ee
The quantities $A$ and $B$ have been known for over a
decade~\cite{NLOTwoJets}, but $C$ has not yet been computed.
\end{enumerate}

\begin{figure}[t]
\begin{minipage}[t]{0.5\linewidth}
\centering\includegraphics[width=2.7truein]{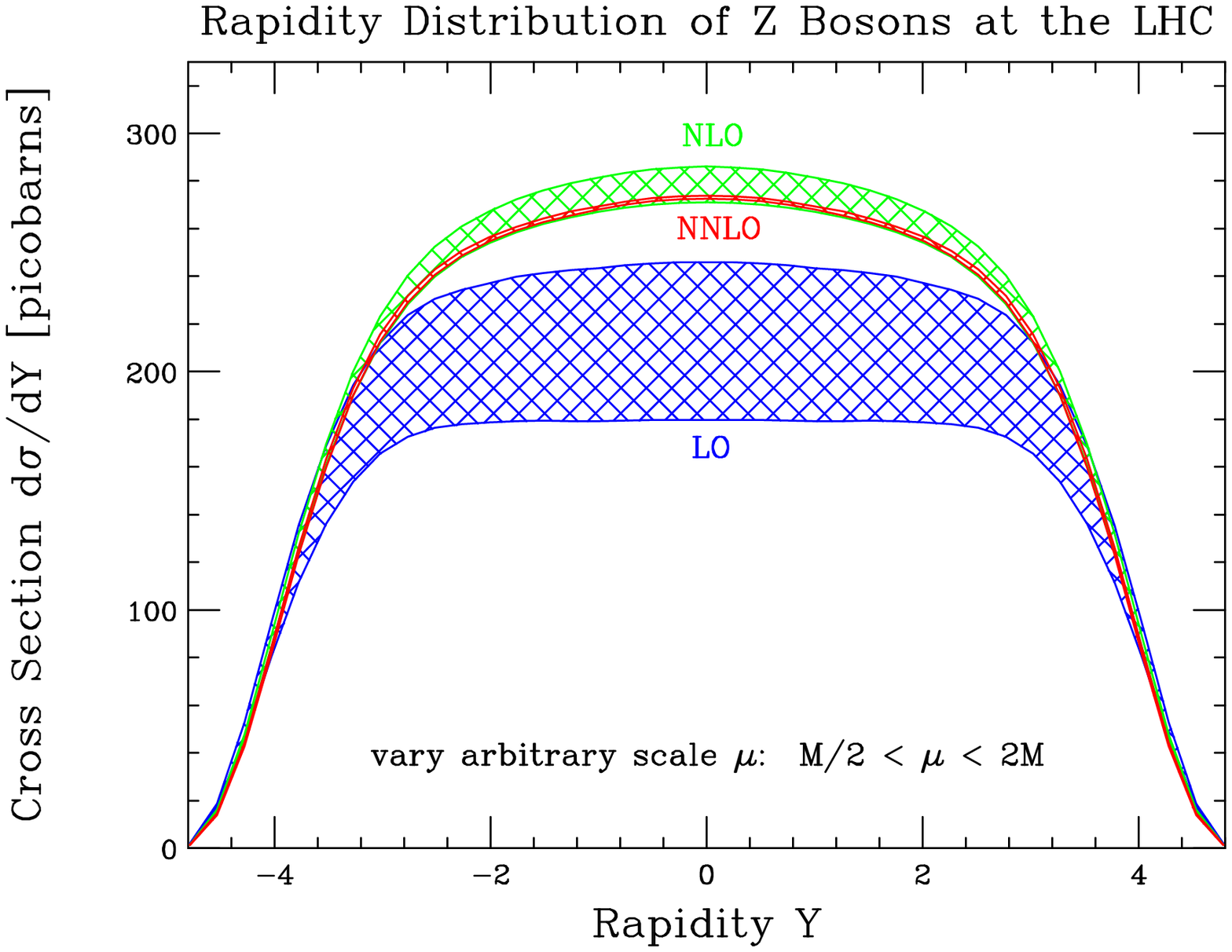} 
\caption[a]{\small LHC $Z$ production~\cite{NNLOWZ}.~~~~~~~~~~
}
\label{LHCZFigure}
\end{minipage}
\begin{minipage}[t]{0.5\linewidth}
\centering\includegraphics[width=2.4truein]{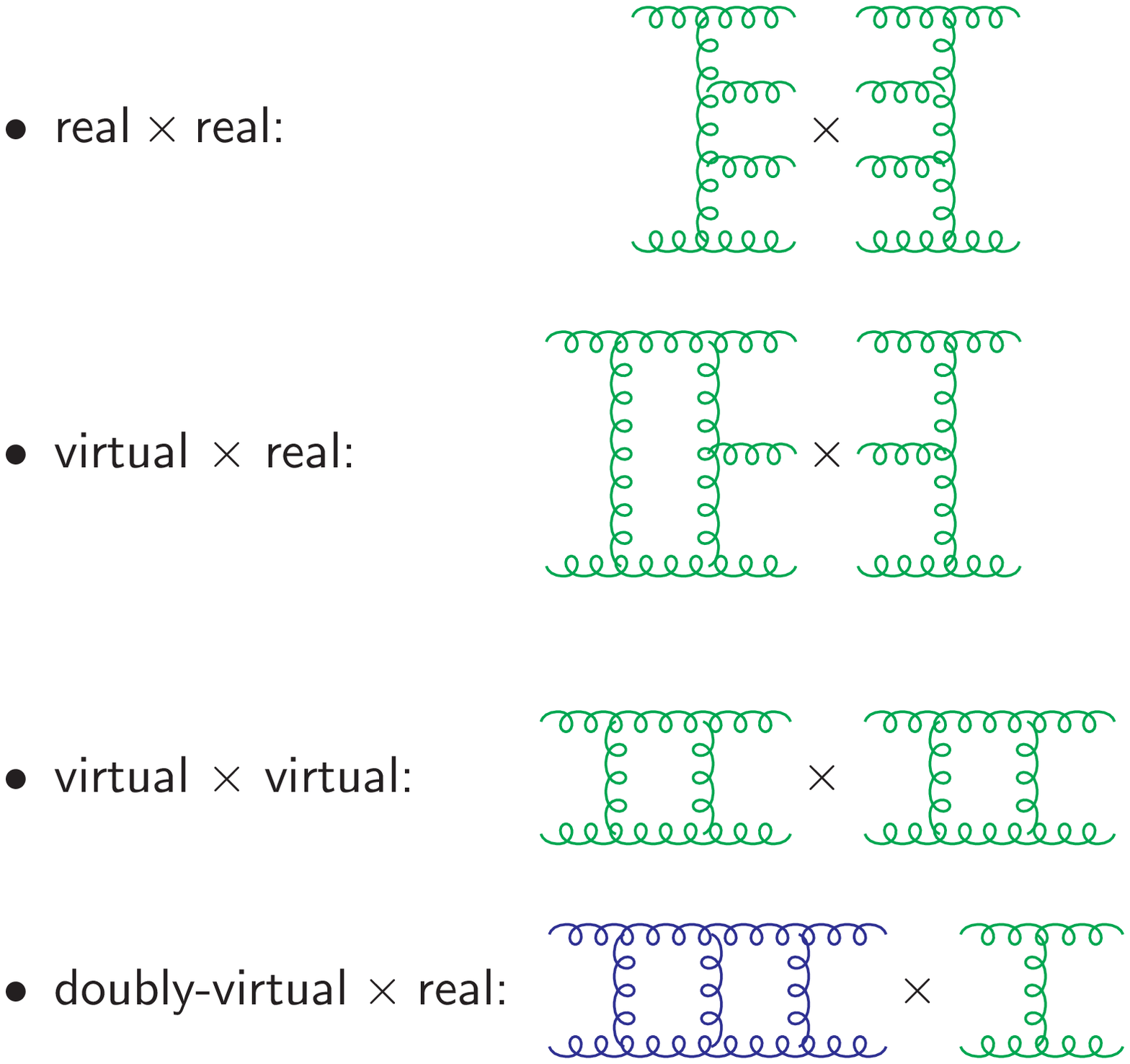}
\caption[a]{\small Purely gluonic contributions \\ 
  to $\hat{\sigma}_{gg \to jjX}$ at NNLO.}
\label{IngredientsFigure}
\end{minipage}
\end{figure}

Indeed, the NNLO terms are unknown for all but a handful of 
collider processes.  Computing a wide range of processes at NNLO
is the goal of a large amount of recent effort in perturbative
QCD~\cite{TGLeptonPhoton}. 
As an example of the improved precision that could result
from this program, consider the production
of a virtual photon, $W$ or $Z$ boson via the Drell-Yan
process at the Tevatron or LHC.
The total cross section for this process was first computed at NNLO in 
1991~\cite{NNLOTOTALDY}.  Last year, the rapidity distribution of 
the vector boson also became available at this order~\cite{NNLODY,NNLOWZ}, 
as shown in \fig{LHCZFigure}.  The rapidity is defined in terms
of the energy $E$ and longitudinal momentum $p_z$ of the vector
boson in the center-of-mass frame, 
$Y \equiv \frac{1}{2} \log \left(\frac{E+p_z}{E-p_z} \right)$.
It determines where the vector boson decays within the detector,
or outside its acceptance.
The rapidity is sensitive to the $x$ values of the incoming partons. 
At leading order in QCD, $x_1 = e^Ym_V/\sqrt{s}$, 
$x_2 = e^{-Y}m_V/\sqrt{s}$, where $m_V$ is the vector boson mass.

The LHC will produce roughly 100 million $W$s and 10 million $Z$s per year
in detectable (leptonic) decay modes.  LHC experiments will be able to
map out the curve in \fig{LHCZFigure} with exquisite precision, and use it to
constrain the parton distributions --- in the same detectors
that are being used to search for new physics in other channels,
often with similar $q\bar{q}$ initial states.  By taking ratios of the other
processes to the ``calibration'' processes of single $W$ and $Z$ production,
many experimental uncertainties, including those associated with the
initial state parton distributions, drop out.  Thus \fig{LHCZFigure}
plays a role as a ``partonic luminosity monitor''~\cite{plum}. 
To get the full benefit of the remarkable experimental precision, though, 
the theory uncertainty must approach the 1\% level.  As seen from the
uncertainty bands in the figure, this precision is only achievable at NNLO.
The bands are estimated by varying the arbitrary renormalization and
factorization scales $\mu_R$ and $\mu_F$ (set to a common value $\mu$)
from $m_V/2$ to $2 m_V$.  A computation to all orders in $\alpha_s$
would have no dependence on $\mu$.  Hence the $\mu$-dependence of a
fixed order computation is related to the size of the missing
higher-order terms in the series.  Although sub-1\% uncertainties
may be special to $W$ and $Z$ production at the LHC, similar
qualitative improvements in precision will be achieved for many 
other processes, such as di-jet production, as the NNLO terms are
completed.

Even within the NNLO terms in the partonic cross section, there are 
several types of ingredients.  This feature is illustrated 
in~\fig{IngredientsFigure} for
the purely gluonic contributions to di-jet production, 
$\hat{\sigma}_{gg \to jjX}$.
In the figure, individual Feynman graphs stand for full amplitudes
interfered ($\times$) with other amplitudes, in order to produce contributions 
to a cross section.  There may be 2, 3, or 4 partons in the final state.
Just as in QED it is impossible to define an outgoing electron with no
accompanying cloud of soft photons, also in QCD sensible observables
require sums over final states with different numbers of partons.
Jets, for example, are defined by a certain amount of energy into
a certain conical region.  At leading order, that energy typically
comes from a single parton, but at NLO there may be two
partons, and at NNLO three partons, within the jet cone.  

Each line in \fig{IngredientsFigure} results in a cross-section 
contribution containing severe infrared divergences, which 
are traditionally regulated by dimensional regulation with $D=4-2\e$.
Note that this regulation breaks the classical conformal invariance
of QCD, and the classical and quantum conformal invariance of 
$\Neqfour$ super-Yang-Mills theory.
Each contribution contains poles in $\e$ ranging from $1/\e^4$ to $1/\e$.
The poles in the real contributions come from regions of phase-space where
the emitted gluons are soft and/or collinear.  The poles in the virtual
contributions come from similar regions of virtual loop integration.
The virtual $\times$ real contribution obviously has a mixture of the two.
The Kinoshita-Lee-Nauenberg theorem~\cite{KLN} guarantees that
the poles all cancel in the sum, for properly-defined, 
short-distance observables, after renormalizing the coupling constant 
and removing initial-state collinear singularities associated with 
renormalization of the pdfs. 

A critical ingredient in any NNLO prediction is the set of two-loop
amplitudes, which enter the doubly-virtual $\times$ real interference
in \fig{IngredientsFigure}.  Such amplitudes require
dimensionally-regulated all-massless two-loop integrals depending on at 
least one dimensionless ratio, which were only computed
beginning in 1999~\cite{Smirnov,Tausk,TwoloopTensorIntegrals}.
They also receive contributions from many Feynman diagrams, with lots 
of gauge-dependent cancellations between them.  It is of interest
to develop more efficient, manifestly gauge-invariant methods 
for combining diagrams, such as the unitarity or cut-based
method successfully applied at one loop~\cite{LoopReview}
and in the initial two-loop computations~\cite{AllPlusTwo}.


\section{$\Neqfour$ Super-Yang-Mills Theory as a Testing Ground for QCD}
\label{TestingGroundSection}

$\Neqfour$ super-Yang-Mills theory serves an excellent testing ground for
perturbative QCD methods.  For $n$-gluon scattering at tree level,
the two theories in fact give identical predictions. (The extra fermions
and scalars of MSYM can only be produced in pairs; hence they only appear
in an $n$-gluon amplitude at loop level.)  Therefore any consequence
of $\Neqfour$ supersymmetry, such as Ward identities among scattering
amplitudes~\cite{SWI}, automatically applies to tree-level gluonic scattering
in QCD~\cite{ParkeTaylorSWI}.  Similarly, at tree level Witten's topological
string~\cite{WittenI} produces MSYM, but implies twistor-space
localization properties for QCD tree amplitudes.  (Amplitudes with
quarks can be related to supersymmetric amplitudes with gluinos using
simple color manipulations.)

\subsection{Pole Structure at One and Two Loops}

\begin{figure}[t]
\centering\includegraphics[width=3.0truein]{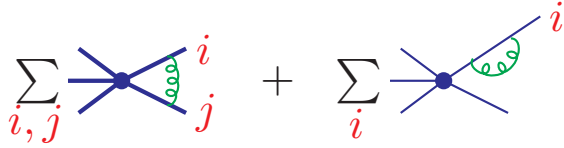}
\caption[a]{\small Illustration of soft-collinear (left) and
 pure-collinear (right) one-loop divergences.}
\label{Catani1Figure}
\end{figure}

At the loop-level, MSYM becomes progressively more removed from QCD.
However, it can still illuminate general properties of scattering
amplitudes, in a calculationally simpler arena.  Consider 
the infrared singularities of one-loop massless gauge theory amplitudes.
In dimensional regularization, the leading singularity is $1/\e^2$,
arising from virtual gluons which are both soft {\it and} collinear 
with respect to a second gluon or another massless particle.
It can be characterized by attaching a gluon to any pair of external legs of
the tree-level amplitude, as in the left graph in 
\fig{Catani1Figure}.  Up to color factors, this leading divergence is the
same for MSYM and QCD.  There are also purely collinear terms associated
with individual external lines, as shown in the right graph in
\fig{Catani1Figure}.  The pure-collinear terms have a simpler form than
the soft terms, because there is less tangling of color indices, but they
do differ from theory to theory.  

The full result for one-loop divergences can be expressed
as an operator $\bI^{(1)}(\e)$ which acts on the color indices of
the tree amplitude~\cite{CataniSingular}.  
Treating the $L$-loop amplitude as a vector in 
color space, $| \cA_n^{(L)} \rangle$, the one-loop result is
\be 
| \cA_n^{(1)} \rangle = \bI^{(1)}(\e) | \cA_n^{(0)} \rangle 
              + | \cA_n^{(1),{\rm fin}} \rangle \,,
\label{CataniOneloop}
\ee
where $| \cA_n^{(1),{\rm fin}} \rangle$ is finite as $\e\to0$, and
\be 
\bI^{(1)}(\e) = {1\over2} { e^{\e \gamma} \over \Gamma(1-\e) }
 \sum_{i=1}^n \sum_{j\neq i}^n \bT_i \cdot \bT_j
 \left[ {1\over\e^2} + {\gamma_i \over \bT_i^2} {1\over\e} \right]
 \left( { \mu_R^2 \over -s_{ij} } \right)^{\e} \,,
\label{I1def}
\ee 
where $\gamma$ is Euler's constant and $s_{ij} = (k_i+k_j)^2$
is a Mandelstam invariant.
The color operator $\bT_i \cdot \bT_j = T_i^a T_j^a$ and
factor of $(\mu_R^2/(-s_{ij}))^{\e}$
arise from soft gluons exchanged between legs $i$ and $j$, as in the left
graph in \fig{Catani1Figure}.  
The pure $1/\e$ poles terms proportional to $\gamma_i$ have been written 
in a symmetric fashion, which slightly obscures the fact that the color 
structure is actually simpler. We can use the equation which represents 
color conservation in the color-space notation, $\sum_{j=1}^n \bT_j = 0$, 
to simplify the result.  At order $1/\e$ we may neglect the 
$(\mu_R^2/(-s_{ij}))^{\e}$ factor in the $\gamma_i$ terms, and we have
$\sum_{j\neq i}^n \bT_i \cdot \bT_j\ \gamma_i/ \bT_i^2
 = - \gamma_i$.  So the color structure of the pure $1/\e$ term is 
actually trivial.  For an $n$-gluon amplitude, the factor $\gamma_i$
is set equal to its value for gluons, which turns out to be 
$\gamma_g = b_0$, the one-loop coefficient in the $\beta$-function.  
Hence the pure-collinear contribution vanishes for MSYM, but not for QCD.

The divergences of two-loop amplitudes can be described in the same
formalism~\cite{CataniSingular}. 
The relation to soft-collinear factorization has been made more 
transparent by Sterman and Tejeda-Yeomans, who also predicted the 
three-loop behavior~\cite{STY}. 
Decompose the two-loop amplitude $| \cA_n^{(2)} \rangle$ as 
\be 
| \cA_n^{(2)} \rangle = \bI^{(2)}(\e) | \cA_n^{(0)} \rangle 
              + \bI^{(1)}(\e) | \cA_n^{(1)} \rangle 
              + | \cA_n^{(2),{\rm fin}} \rangle \,,
\label{CataniTwoloop}
\ee
where $| \cA_n^{(2),{\rm fin}} \rangle$ is finite as $\e\to0$ and
\bea 
\bI^{(2)}(\e) &=& - {1\over2} \bI^{(1)}(\e)
  \left( \bI^{(1)}(\e) + {2 b_0 \over \e } \right)
  + { e^{-\e\gamma} \Gamma(1-2\e) \over \Gamma(1-\e) }
  \left( { b_0 \over \e } + K \right) \bI^{(1)}(2\e) 
\nonumber \\
&& \hskip0.1cm 
+ { e^{\e\gamma} \over 4\e \, \Gamma(1-\e) } 
  \biggl[ 
  - \sum_{i=1}^n \sum_{j\neq i}^n \, \bT_i \cdot \bT_j
     { H_i^{(2)} \over \bT_i^2 } \, 
    \Bigl( { \mu^2 \over -s_{ij} } \Bigr)^{2\e}
       + \hat\bH^{(2)} \biggr] \,.
\label{I2def}
\eea 
Here $K$ and $H_i^{(2)}$ depend on the theory, and $H_i^{(2)}$, like
$\gamma_i$, also depends on the external leg $i$.  For QCD, $K$ has long
been known from soft-gluon resummation~\cite{KodairaTrentadue}, while
$H_i^{(2)}$ were found by explicit computation of four-parton two-loop
scattering amplitudes~\cite{Twoloopqqqqqqgg,Twoloopgggg,Twoloopgggghel}.
For MSYM, the quantities are naturally simpler, 
\bea
 K^{\Neqfour} &=& - \zeta_2 \, C_A 
   \,,  
  \label{CataniKNeq4}\\
 H_i^{(2), \, \Neqfour} &=& {\zeta_3\over2}  C_A^2 \,,
\label{HNeq4}
\eea
where $C_A = \Nc$ is the adjoint Casimir value.
The quantity $\hat\bH^{(2)}$ has non-trivial, but
purely subleading-in-$\Nc$, color structure.
It is associated with soft, rather than collinear,
momenta~\cite{Twoloopgggghel,STY}, so it is theory-independent, up to
color factors. An ansatz for it for general $n$ has been presented
recently~\cite{Splitggg}.

\subsection{Recycling Cuts in MSYM}

\begin{figure}[t]
\centering\includegraphics[width=3.5truein]{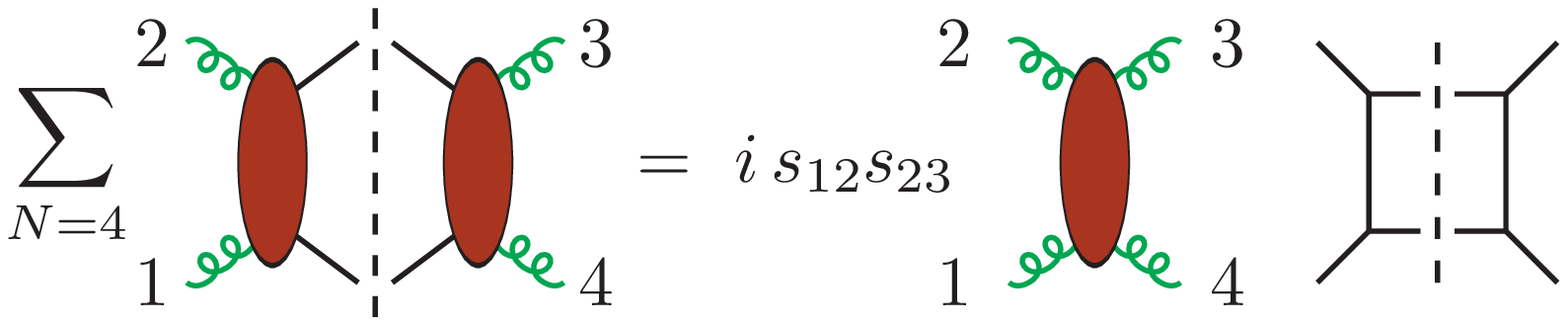}
\caption[a]{\small The one-loop two-particle cuts for the four-point
amplitude in MSYM reduce to the tree amplitude multiplied by 
a cut scalar box integral (for any set of four external states).}
\label{Neq4cutFigure}
\end{figure}

An efficient way to compute loop amplitudes, particularly in theories 
with a great deal of supersymmetry, is to use unitarity and reconstruct 
the amplitude from its cuts~\cite{LoopReview,Splitggg}.
For the four-gluon amplitude in MSYM, the two-loop structure, and much of 
the higher-loop structure, follows from a simple property of the
one-loop two-particle cut in this theory.  For simplicity, we strip the
color indices off of the four-point amplitude $\cA_4^{\tree}$, by 
decomposing it into color-ordered amplitudes $A_4^{\tree}$, whose 
coefficients are traces of $\SUNc$ generator matrices (Chan-Paton factors),
\bea 
\cA_4^{\tree}(k_1,a_1;k_2,a_2;k_3,a_3;k_4,a_4) &=& g^{2} 
 \sum_{\rho\in S_4/Z_4} \Tr( T^{a_{\rho(1)}} T^{a_{\rho(2)}} 
   T^{a_{\rho(3)}} T^{a_{\rho(4)}} )  
 \nonumber \\ && \hskip1.3cm \times
 A_4^{\tree}(k_{\rho(1)}, k_{\rho(2)}, k_{\rho(3)}, k_{\rho(4)})\,.
\label{Fourtreecolor}
\eea
The two-particle cut can be written as a product of two four-point 
color-ordered amplitudes, summed over the pair of intermediate $\Neqfour$
states $S,S'$ crossing the cut, which evaluates to
\bea
&&\sum_{S,S' \in \Neqfour} A_4^\tree(k_1,k_2,\ell_S,-\ell_{S'}') 
   \times A_4^\tree(\ell_{S'}',-\ell_S,k_3,k_4)
\nonumber \\ && \hskip0.2cm
 = i s_{12} s_{23} A_4^\tree(k_1,k_2,k_3,k_4) 
     \times { 1 \over (\ell' - k_1)^2 } { 1 \over (\ell - k_3)^2 } 
\,,
\label{Neq4cutEqn}
\eea
where $\ell' = \ell - k_1 - k_2$. 
This equation is also shown in \fig{Neq4cutFigure}. 
The scalar propagator factors
in \eqn{Neq4cutEqn} are depicted as solid vertical lines in the figure.
The dashed line indicates the cut.  Thus the cut reduces to the cut of a 
scalar box integral, defined by
\be
\mathcal{I}_4^{D=4-2\e} \equiv \int { d^{4-2\e}\ell \over (2\pi)^{4-2\e} }
      { 1 \over \ell^2 (\ell-k_1)^2 (\ell-k_1-k_2)^2 (\ell+k_4)^2 } \,.
\label{BoxDef}
\ee
One of the virtues of \eqn{Neq4cutEqn} is that it is valid for arbitrary
external states in the $\Neqfour$ multiplet, although only external gluons
are shown in \fig{Neq4cutFigure}.  Therefore it can be re-used at higher
loop order, for example by attaching yet another tree to the left.

\begin{figure}[t]
\centering\includegraphics[width=5.0truein]{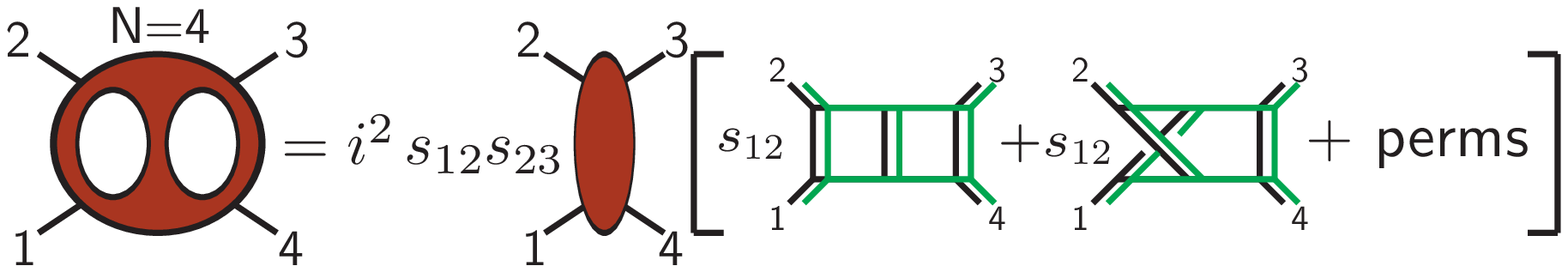}
\caption[a]{\small The two-loop $\ggtogg$ amplitude in MSYM~\cite{BRY,BDDPR}.  
The blob on the right represents the color-ordered tree amplitude 
$A_4^\tree$. (The quantity $s_{12} s_{23} A_4^\tree$
transforms symmetrically under gluon interchange.)
In the the brackets, black lines are kinematic $1/p^2$ propagators, 
with scalar ($\phi^3$) vertices. 
Green lines are color $\delta^{ab}$ propagators, with structure constant
($f^{abc}$) vertices.  The permutation sum is over the three cyclic
permutations of legs 2,3,4, and makes the amplitude Bose symmetric.}
\label{Neq42loopFigure}
\end{figure}

At two loops, the simplicity of \eqn{Neq4cutEqn} made it possible 
to compute the two-loop $\ggtogg$ scattering amplitude in that 
theory (in terms of specific loop integrals) in 1997~\cite{BRY}, 
four years before the analogous computations in
QCD~\cite{Twoloopgggg,Twoloopgggghel}.
All of the loop momenta in the numerators of the Feynman diagrams can
be factored out, and only two independent loop integrals appear, the
planar and nonplanar scalar double box integrals.
The result can be written in an appealing diagrammatic form, 
\fig{Neq42loopFigure}, where the color algebra has the same form as 
the kinematics of the loop integrals~\cite{BDDPR}.  

\begin{figure}[t]
\begin{minipage}[t]{0.5\linewidth}
\centering\includegraphics[width=2.5truein]{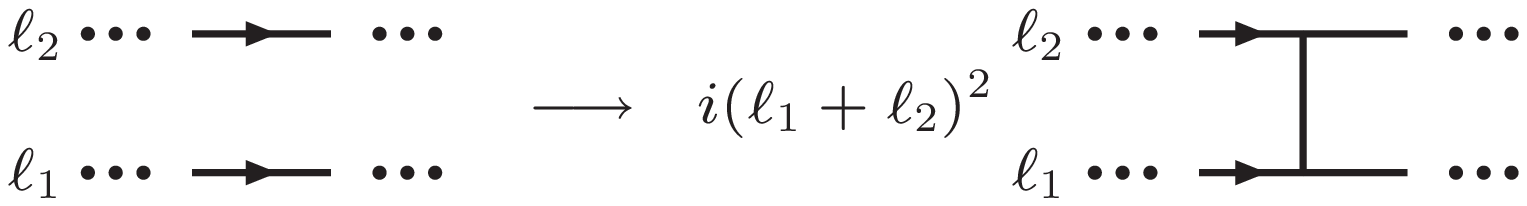}
\caption[a]{\small The rung rule for MSYM.~~~~~}
\label{RungRuleFigure}
\end{minipage}
\begin{minipage}[t]{0.5\linewidth}
\centering\includegraphics[width=2.5truein]{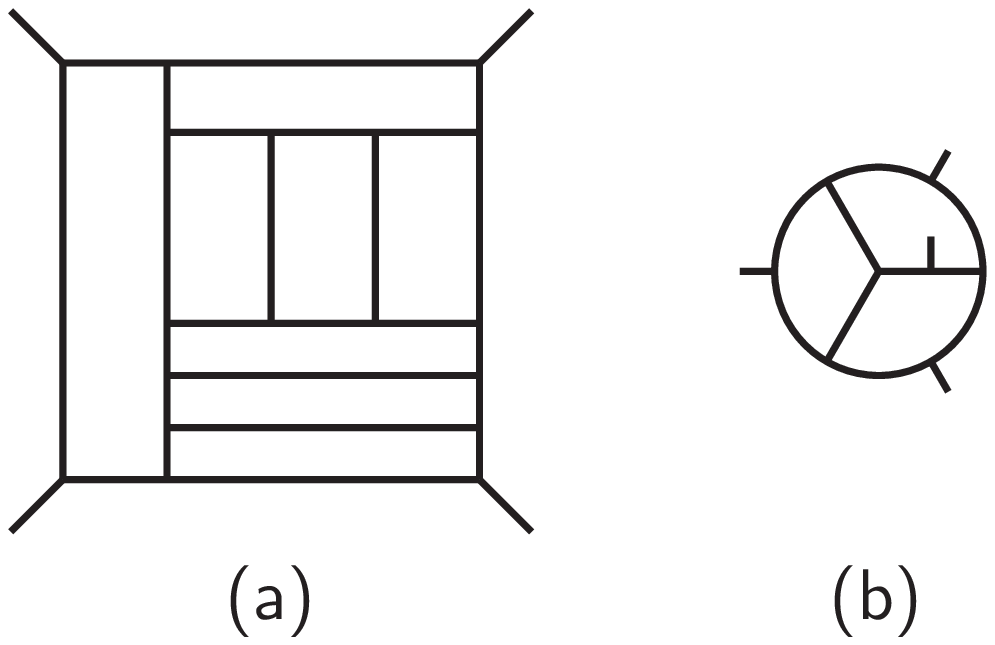}
\caption[a]{\small (a) Example of a `Mondrian' diagram which can
 be determined recursively from the rung rule.  (b) The first
 non-vanishing, non-Mondrian diagrams appear at three loops in nonplanar, 
 subleading-color contributions.}
\label{MonnonMonFigure}
\end{minipage}
\end{figure}

At higher loops, \eqn{Neq4cutEqn} leads to a ``rung rule''~\cite{BRY}
for generating a class of $(L+1)$-loop contributions from $L$-loop contributions.
The rule states that one can insert into a $L$-loop contribution 
a rung, {\it i.e.} a scalar propagator, transverse to two parallel 
lines carrying momentum $\ell_1+\ell_2$, along with a factor of
$i(\ell_1+\ell_2)^2$ in the numerator, as shown in \fig{RungRuleFigure}.
Using this rule, one can construct recursively the external
and loop-momentum-containing numerators factors associated with
every $\phi^3$-type diagram that can be reduced to trees by a sequence
of two-particle cuts, such as the diagram in \fig{MonnonMonFigure}{\it a}.
Such diagrams can be termed ``iterated 2-particle cut-constructible,''
although a more compact notation might be `Mondrian' diagrams,
given their resemblance to Mondrian's paintings.
Not all diagrams can be computed in this way.
The diagram in \fig{MonnonMonFigure}{\it b} is not in the `Mondrian'
class, so it cannot be determined from two-particle cuts.  Instead, 
evaluation of the three-particle cuts shows that it appears with 
a non-vanishing coefficient in the subleading-color contributions to the 
three-loop MSYM amplitude.

\section{Iterative Relation in $\Neqfour$ Super-Yang-Mills Theory}
\label{IterativeSection}

Although the two-loop $\ggtogg$ amplitude in MSYM was expressed
in terms of scalar integrals in 1997~\cite{BRY}, and the integrals
themselves were computed as a Laurent expansion about $D=4$ 
in 1999~\cite{Smirnov,Tausk}, the expansion of the $\Neqfour$ amplitude
was not inspected until last fall~\cite{PlanarMSYM}, considerably
after similar investigations for QCD and $\Neqone$ super-Yang-Mills
theory~\cite{Twoloopgggg,Twoloopgggghel}.  It was found to have
a quite interesting ``iterative'' relation, when expressed in terms of 
the one-loop amplitude and its square.

\begin{figure}[t]
\centering\includegraphics[width=4.8truein]{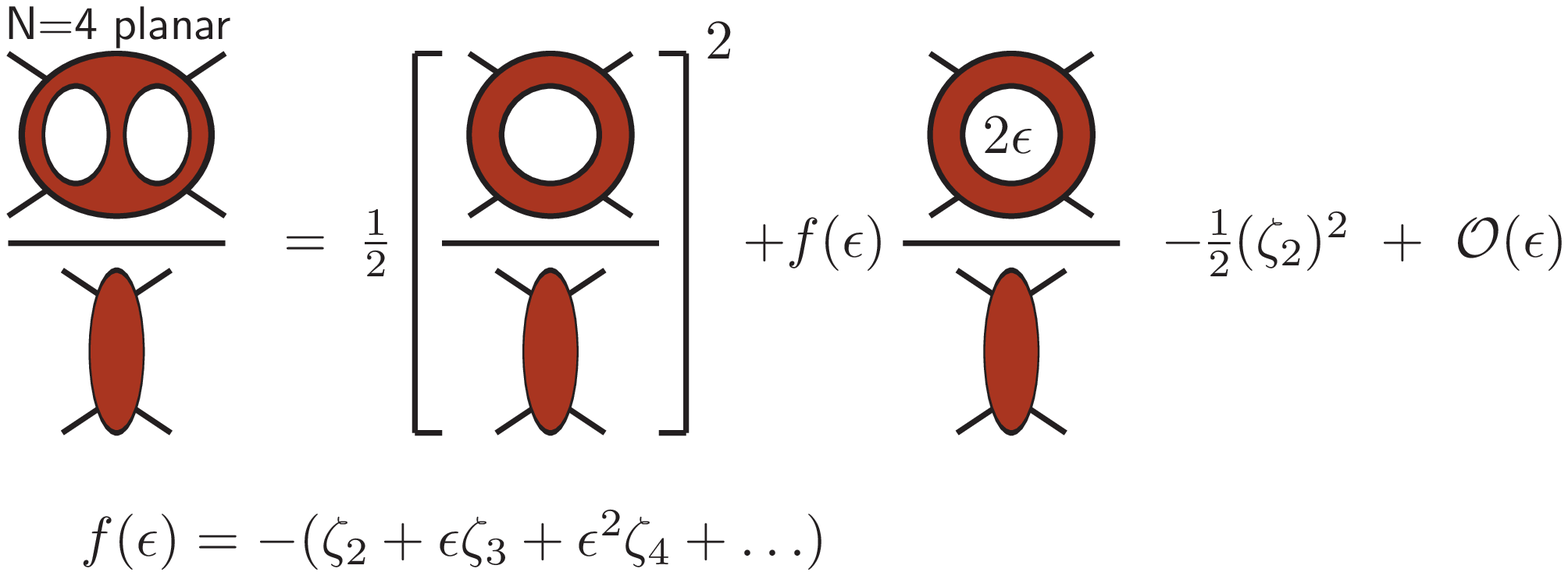}
\caption[a]{\small Schematic depiction of the iterative
relation~(\ref{TwoloopOneloop}) between two-loop and one-loop MSYM
amplitudes.}
\label{Neq42iterateFigure}
\end{figure}

At leading color, the $L$-loop $\ggtogg$ amplitude has the same
single-trace color decomposition as the tree amplitude, \eqn{Fourtreecolor}.
Let $M_4^{(L)}$ be the ratio of this leading-color, color-ordered
amplitude to the corresponding tree amplitude, omitting also several
conventional factors, 
\be
A_4^{(L),\Neqfour\ {\rm planar}} = 
\biggl[ { 2 e^{- \e \gamma} g^2 \Nc \over (4\pi)^{2-\e} } \biggr]^{L}
A_4^\tree \times M_4^{(L)} \,.
\label{M4def}
\ee 
Then the iterative relation (see also \fig{Neq42iterateFigure}) is
\be
M_4^{\twoloop}(\e) =  {1 \over 2} \Bigl(M_4^{\oneloop}(\e) \Bigr)^2
             + f(\e) \, M_4^{\oneloop}(2\e) 
             - {1\over2} (\zeta_2)^2\ +\ \Ord(\e) \,,
\label{TwoloopOneloop}
\ee
where $f(\e) \equiv (\psi(1-\e)-\psi(1))/\e
= - (\zeta_2 + \zeta_3 \e + \zeta_4 \e^2 + \cdots)$.

The analogous relation for gluon-gluon scattering at two loops in QCD 
takes a similar form at the level of the pole terms in $\e$, thanks to the
general result~(\ref{CataniTwoloop}).  But the finite remainder 
$-{1\over2}(\zeta_2)^2$ is replaced by approximately six pages of formulas
(!), including a plethora of polylogarithms, logarithms and polynomials 
in ratios of invariants $s/t$, $s/u$ and $t/u$~\cite{Twoloopgggghel}.
The polylogarithm is defined by
\be
\Li_m(x) = \sum_{i=1}^\infty { x^i \over i^m } 
         = \int_0^x {dt \over t} \Li_{m-1}(t),
\qquad \Li_1(x) = - \ln(1-x).
\label{Lidef}
\ee
It appears with degree $m$ up to 4 at the finite, order $\e^0$, level;
and up to degree $4-i$ in the $\Ord(\e^{-i})$ terms.
In the case of MSYM, identities relating these polylogarithms are 
needed to establish \eqn{TwoloopOneloop}.

Although the $\Ord(\e^0)$ term in \eqn{TwoloopOneloop} is miraculously
simple, as noted above the behavior of the pole terms is not a miracle.
It is dictated in general terms by the cancellation of infrared
divergences between virtual corrections and real emission~\cite{KLN}.
Roughly speaking, for this cancellation to take place, the virtual
terms must resemble lower-loop amplitudes, and the real terms must
resemble lower-point amplitudes, in the soft and collinear regions
of loop or phase-space integration.

At the level of the finite terms, the iterative
relation~(\ref{TwoloopOneloop}) can be understood in the Regge/BFKL 
limit where $s \gg t$, because it then corresponds to exponentiation 
of large logarithms of $s/t$~\cite{BFKL}.  For general values of $s/t$, 
however, there is no such argument.

The relation is special to $D=4$, where the theory is conformally
invariant.  That is, the $\Ord(\e^1)$
remainder terms cannot be simplified significantly.  For example,
the two-loop amplitude $M_4^\twoloop(\e)$ contains at $\Ord(\e^1)$
all three independent $\Li_5$ functions, $\Li_5(-s/u)$, $\Li_5(-t/u)$ 
and $\Li_5(-s/t)$, yet $[M_4^{\oneloop}(\e)]^2$ has only the first two 
of these~\cite{PlanarMSYM}.

The relation is also special to the planar, leading-color limit.
The subleading color-components of the finite remainder 
$| \cA_n^{(2),{\rm fin}} \rangle$ defined by \eqn{CataniTwoloop}
show no significant simplification at all.

For planar amplitudes in the $D \to 4$ limit, however, there is evidence 
that an identical relation also holds for an arbitrary number $n$ of 
external legs, at least for certain ``maximally helicity-violating'' 
(MHV) helicity amplitudes.  This evidence comes from studying the limits of
two-loop amplitudes as two of the $n$ gluon momenta become
collinear~\cite{PlanarMSYM,Splitggg,SplitBG}.
(Indeed, it was by analyzing these limits that the relation for $n=4$
was first uncovered.)  The collinear limits turn out to be consistent 
with the same \eqn{TwoloopOneloop} with $M_4$ replaced by $M_n$
everywhere~\cite{PlanarMSYM}, {\it i.e.}
\be
M_n^{\twoloop}(\e) =  {1 \over 2} \Bigl(M_n^{\oneloop}(\e) \Bigr)^2
             + f(\e) \, M_n^{\oneloop}(2\e) 
             - {1\over2} (\zeta_2)^2\ +\ \Ord(\e) \,.
\label{TwoloopOneloopn}
\ee
The collinear consistency does not constitute a proof of
\eqn{TwoloopOneloopn}, but in light of the remarkable properties of MSYM, 
it would be surprising if it were not true in the MHV case.
Because the direct computation of two-loop amplitudes for $n>4$ seems 
rather difficult, it would be quite interesting to try to examine 
the twistor-space properties of \eqn{TwoloopOneloopn}, along the lines of
refs.~\cite{WittenI,CSWII}.
(The right-hand-side of~\eqn{TwoloopOneloopn} is
not completely specified at order $1/\e$ and $\e^0$ for $n>4$.  
The reason is that the order $\e$ and $\e^2$ terms
in $M_n^{\oneloop}(\e)$, which contribute to the first term in 
\eqn{TwoloopOneloopn} at order $1/\e$ and $\e^0$, 
contain the $D=6-2\e$ pentagon integral~\cite{SelfDual}, 
which is not known in closed form.  
On the other hand, the differential equations this integral satisfies 
may suffice to test the twistor-space behavior. Or one may examine
just the finite remainder $M_n^{(L),{\rm fin}}$ defined
via~\eqn{CataniTwoloop}.)

\begin{figure}[t]
\centering\includegraphics[width=4.8truein]{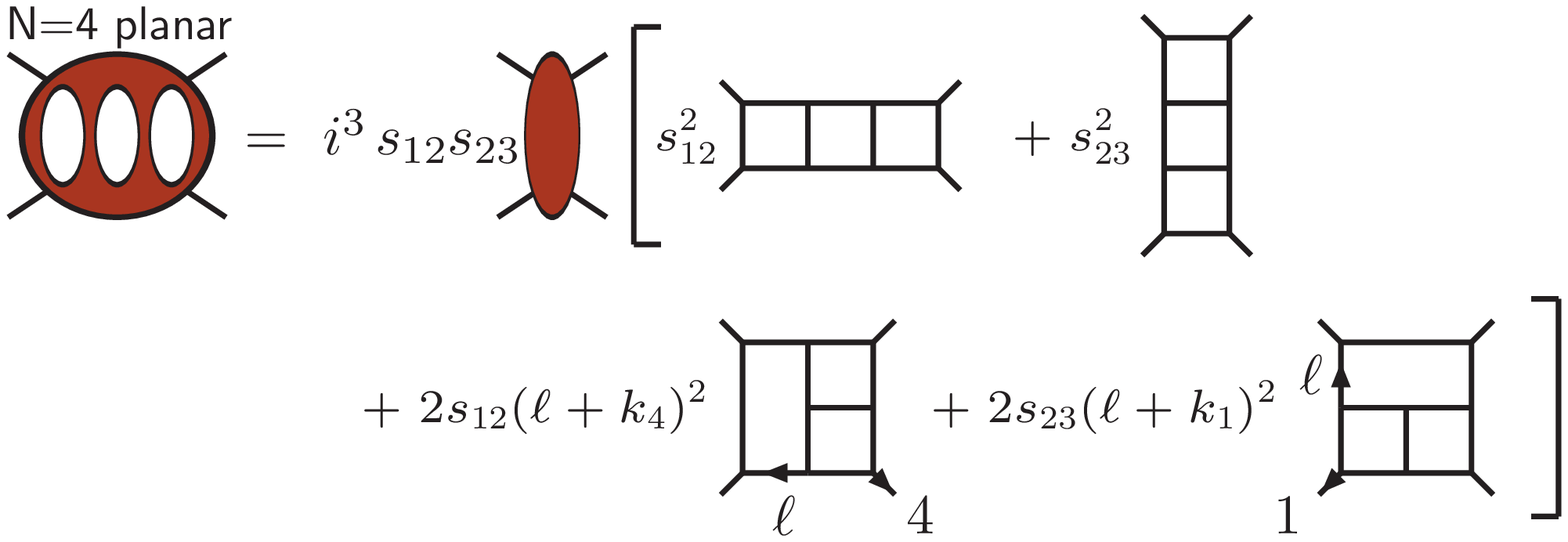}
\caption[a]{\small Graphical representation of the three-loop
amplitude for MSYM in the planar limit.}
\label{Neq43planarFigure}
\end{figure}

It may soon be possible to test whether an iterative relation for planar
MSYM amplitudes extends to three loops.  An ansatz for the three-loop
planar $gg\to gg$ amplitude, shown in \fig{Neq43planarFigure}, was
provided at the same time as the two-loop result, in 1997~\cite{BRY}.
The ansatz is based on the ``rung-rule'' evaluation of 
the iterated 2-particle cuts, plus the 3-particle cuts 
with intermediate states in $D=4$;
the 4-particle cuts have not yet been verified. 
Two integrals, each beginning at $\Ord(\e^{-6})$, 
are required to evaluate the ansatz in a Laurent expansion
about $D=4$.  (The other two integrals are related by $s\lr t$.)
The triple ladder integral on the top line of
\fig{Neq43planarFigure} was evaluated last year by Smirnov, all the way
through $\Ord(\e^0)$~\cite{SmirnovTripleLadder}.
Evaluation of the remaining integral, which contains a factor of 
$(\ell+k_4)^2$ in the numerator, is in progress~\cite{SmirnovExtraIntegral};
all the terms through $\Ord(\e^{-2})$ agree with predictions~\cite{STY}, 
up to a couple of minor corrections.

\section{Significance of Iterative Behavior?}
\label{SigSection}

It is not yet entirely clear why the two-loop four-point amplitude,
and probably also the $n$-point amplitudes, have the iterative
structure~(\ref{TwoloopOneloopn}).  However, one can speculate that
it is from the need for the perturbative series to be summable into
something which becomes ``simple'' in the planar strong-coupling limit,
since that corresponds, via AdS/CFT, to a weakly-coupled supergravity
theory.  The fact that the relation is special to the conformal limit
$D\to4$, and to the planar limit, backs up this speculation.
Obviously it would be nice to have some more information at three
loops.  There have been other hints of an iterative structure in the
four-point correlation functions of chiral primary (BPS) composite
operators~\cite{Edenetal}, but here also the exact structure is not yet
clear.  Integrability has played a key role in recent higher-loop computations
of non-BPS spin-chain anomalous dimensions~\cite{MZ,BKS,BFST,BDS}.
By imposing regularity of the BMN `continuum' limit~\cite{BMN}, a piece of the 
anomalous dimension matrix has even been summed to all orders in $g^2 \Nc$
in terms of hypergeometric functions~\cite{RT}.
The quantities we considered here --- gauge-invariant, but dimensionally
regularized, scattering amplitudes of color non-singlet states ---
are quite different from the composite color-singlet operators usually
treated.  Yet there should be some underlying connection between
the different perturbative series.


\section{Aside:  Anomalous Dimensions in QCD and MSYM}
\label{AnomDimSection}

As mentioned previously, the set of anomalous dimensions for 
leading-twist operators was recently computed at NNLO in
QCD, as the culmination of a multi-year effort~\cite{MVVNNLO}
which is central to performing precise computations of hadron collider
cross sections.  Shortly after the Moch, Vermaseren and Vogt computation,
the anomalous dimensions in MSYM were extracted from this result
by Kotikov, Lipatov, Onishchenko and Velizhanin~\cite{KLOV}.
(The MSYM anomalous dimensions are universal; supersymmetry implies
that there is only one independent one for each Mellin moment $j$.)
This extraction was non-trivial, because MSYM contains scalars,
interacting through both gauge and Yukawa interactions, whereas QCD does
not.  However, Kotikov {\it et al.} noticed, from comparing NLO
computations in both leading-twist anomalous dimensions and BFKL
evolution, that the ``most complicated terms'' in the QCD computation
always coincide with the MSYM result, once the gauge group representation
of the fermions is shifted from the fundamental to the adjoint
representation.  One can define the ``most complicated terms''
in the $x$-space representation of the anomalous dimensions --- {\it i.e.}
the splitting kernels --- as follows:
Assign a logarithm or factor of $\pi$
a transcendentality of 1, and a polylogarithm $\Li_m$ or factor of
$\zeta_m = \Li_m(1)$ a transcendentality of $m$.  
Then the most complicated terms are those with leading transcendentality.
For the NNLO anomalous dimensions, this turns out to be transcendentality 4.  
(This rule for extracting the MSYM terms from QCD has also been found
to hold directly at NNLO, for the doubly-virtual contributions~\cite{Splitggg}.)
Strikingly, the NNLO MSYM anomalous dimension obtained for $j=4$ by this
procedure agrees with a previous result derived by assuming 
an integrable structure for the planar three-loop contribution to the 
dilatation operator~\cite{BKS}.


\section{Conclusions and Outlook}
\label{ConclusionSection}

$\Neqfour$ super-Yang-Mills theory is an excellent testing ground
for techniques for computing, and understanding the structure of,
QCD scattering amplitudes which are needed for precise theoretical 
predictions at high-energy colliders.  One can even learn something 
about the structure of $\Neqfour$ super-Yang-Mills theory in the process, 
although clearly there is much more to be understood.  
Some open questions include:
Is there any AdS/CFT ``dictionary'' for color non-singlet states, like
plane-wave gluons?  Can one recover composite operator correlation
functions from any limits of multi-point scattering amplitudes?
Is there a better way to infrared regulate $\Neqfour$ supersymmetric 
scattering amplitudes, that might be more convenient for approaching 
the AdS/CFT correspondence, such as compactification on a three-sphere,
use of twistor-space, or use of coherent external states?
Further investigations of this arena will surely be fruitful.



\section*{Acknowledgements}
We are grateful to the organizers of {\it Strings04} for putting together
such a stimulating meeting.  This research was supported by the
US Department of Energy under contracts DE-FG03-91ER40662 (Z.B.) 
and DE-AC02-76SF00515  
(L.J.D.), and by the {\it Direction des Sciences de la Mati\`ere\/}
of the {\it Commissariat \`a l'Energie Atomique\/} of France (D.A.K.).


\end{document}